\documentclass[10pt,journal]{IEEEtran} 
\IEEEoverridecommandlockouts
\renewcommand\IEEEkeywordsname{Keywords}
\usepackage{graphicx}

\usepackage{amsmath}
\usepackage{float}
\usepackage{amssymb}
\usepackage{amsfonts}
\usepackage{mathtools}
\usepackage{MnSymbol}
\usepackage{subcaption}
\usepackage{amsthm}
\usepackage[frozencache=true,cachedir=minted-cache]{minted}

\newtheorem{example}{Example}
\newtheorem{remark}{Remark}

\DeclareMathOperator{\luntil}{\mathbf{\mathcal{U}}}
\newcommand{\levent}{\lozenge}
\newcommand{\lalways}{\square}

\makeatletter
\RequirePackage[bookmarks,unicode,colorlinks=true]{hyperref}%
\def\@citecolor{blue}%
\def\@urlcolor{blue}%
\def\@linkcolor{blue}%

\newcommand{\tool}{\texttt{PyTeLo}}
\title{ \LARGE \bf A Flexible and Efficient Temporal Logic \\ Tool for Python: \href{https://github.com/erl-lehigh/PyTeLo}{\texttt{PyTeLo}}}

\author{ Gustavo A. Cardona, Kevin Leahy, Makai Mann, and Cristian-Ioan Vasile
\thanks{Gustavo A. Cardona and Cristian-Ioan Vasile  are with the Mechanical Engineering and Mechanics Department at Lehigh University, PA, USA:$\{$\texttt{gcardona, cvasile\}@lehigh.edu}}
\thanks{Kevin Leahy and Makai Mann are with Massachusetts Institute of Technology Lincoln Laboratory, Lexington, MA 02421, USA 
$\{$\texttt{kevin.leahy, makai.mann\}@ll.mit.edu}}
\thanks{DISTRIBUTION STATEMENT A. Approved for public release. Distribution is unlimited. This material is based upon work supported by the Under Secretary of Defense for Research and Engineering under Air Force Contract No. FA8702-15-D-0001. Any opinions, findings, conclusions or recommendations expressed in this material are those of the author(s) and do not necessarily reflect the views of the Under Secretary of Defense for Research and Engineering. © 2023 Massachusetts Institute of Technology. Delivered to the U.S. Government with Unlimited Rights, as defined in DFARS Part 252.227-7013 or 7014 (Feb 2014). Notwithstanding any copyright notice, U.S. Government rights in this work are defined by DFARS 252.227-7013 or DFARS 252.227-7014 as detailed above. Use of this work other than as specifically authorized by the U.S. Government may violate any copyrights that exist in this work.}
}     
\begin{document}
\maketitle
\begin{abstract}
Temporal logic is an important tool for specifying complex behaviors of systems. It can be used to define properties for verification and monitoring, as well as goals for synthesis tools, allowing users to specify rich missions and tasks.
Some of the most popular temporal logics include Metric Temporal Logic (MTL), Signal Temporal Logic (STL), and weighted STL (wSTL), which also allow the definition of timing constraints.
In this work, we introduce \tool, a modular and versatile Python-based software that facilitates working with temporal logic languages, specifically MTL, STL, and wSTL.
Applying \tool\, requires only a string representation of the temporal logic specification and, optionally, the system's dynamics of interest.
Next, \tool\, reads the specification using an ANTLR-generated parser and generates an Abstract Syntax Tree (AST) that captures the structure of the formula.
For synthesis, the AST serves to recursively encode the specification into a Mixed Integer Linear Program (MILP) that is solved using a commercial solver such as Gurobi.
We describe the architecture and capabilities of \tool\, and provide example applications highlighting its adaptability and extensibility for various research problems.
\end{abstract}


\section{Introduction}

Owing to their rich expressivity, temporal logic formalisms have
proved to be useful in capturing complex and precise high-level mission specifications with logical and temporal constraints~\cite{bellini2000temporal, kress2018synthesis,sun2022multi}. 
Temporal logic formalisms have been widely used in the literature for verifying, synthesizing, and providing guarantees on system behavior.
In this work, we introduce \tool, a Python-based software that facilitates the use of temporal logic languages such as Signal Temporal Logic (STL)~\cite{maler2004}, Weighted Signal Temporal Logic (wSTL)~\cite{mehdipour2020specifying}, and Metric Temporal Logic (MTL)~\cite{brzoska1998programming}.
There are several other tools for working with STL in Python. While \tool\, is general-purpose and allows a user to perform synthesis tasks, many other tools are designed primarily for falsification~\cite{cralley2020,thibeault2021}, evaluation and monitoring~\cite{ulus2019,pyMTL}, or validation~\cite{stl_inspect}, and cannot perform synthesis directly. 
While amenable to synthesis, some are custom-tailored for specific applications, such as machine learning~\cite{Leung2020}, and lack the flexibility of \tool. 
The most closely related tool is \texttt{stlpy}~\cite{kurtz2022}, which can perform evaluation and synthesis, including optimizing a dynamical system using a  Mixed-Integer Linear Program (MILP) formulation. 
\tool\, has these features in addition to a modular design based around the Another Tool for Language Recognition (ANTLR)~\cite{parr2013definitive} framework, which makes it inherently more extensible and flexible. 
Additionally, these tools are not designed to extend the language. 
Most of them work exclusively with either STL~\cite{cralley2020,thibeault2021,stl_inspect,Leung2020,kurtz2022} or MTL~\cite{ulus2019}. 
On the other hand, \tool\, is designed to be extensible to novel syntax and semantics. 
For example, it can easily extend STL to wSTL, allowing users to define preferences and importance over the specification by allocating weights to the operators.
\section{Tool Design}
 In this section, we describe the design of \tool\, (available at \href{https://github.com/erl-lehigh/PyTeLo}{https://github.com/erl-lehigh/PyTeLo}). It is built to efficiently encode languages such as STL, MTL, and wSTL to an optimization problem for finding satisfying traces.
\tool\, is a software tool designed in and for Python that leverages off-the-shelf software such as ANTLR for parsing and creating grammar files and Gurobi~\cite{gurobi} as the optimization solver.
The main objective of \tool\, is to be flexible, modular, and extensible software capable of capturing temporal logic specifications and the dynamics of any appropriate field, e.g., robotics~\cite{jing2016end}, signal processing~\cite{donze2012temporal}, circuits~\cite{bochmann1982hardware}, and biology~\cite{rizk2008continuous}.

\subsection{Architecture \tool} 
\begin{figure*}[t]
    \centering
    \includegraphics[width=\linewidth]{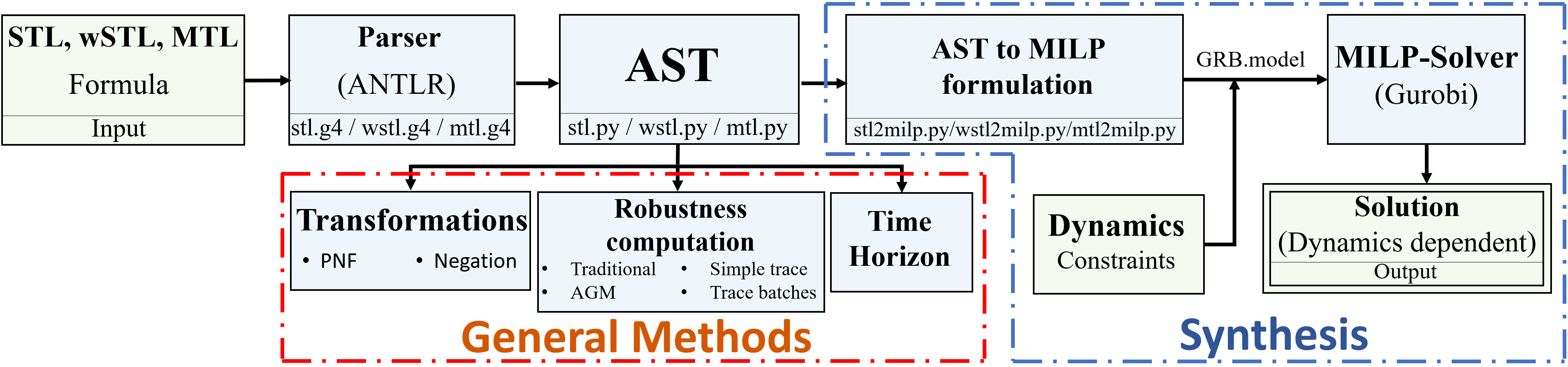}
    \caption{Architecture diagram: green block corresponds to inputs provided by the user and output, and cyan blocks correspond to components of \tool.}
    \label{fig:diagram}
    \vspace{-15pt}
\end{figure*}

The diagram in Fig. \ref{fig:diagram} displays the overall architecture of \tool. 
The inputs and output blocks are green, and the intermediate modules are blue.
The inputs include the temporal logic specification as a string and the dynamics module, which the user can customize according to the application problem.
Intermediate blocks in the diagram consider parsing the specification, creating an AST from it, and two different but complementary groups, \emph{synthesis} (blue dotted box) and \emph{general methods} (red dotted box).
Synthesis contains a translation module to go from AST into a MILP and uses a solver.
General methods leverage the AST structure to compute methods such as transformations to Positive Normal Form (PNF), the negation of the specification, time horizon, and robustness. 

\subsection{Specification formula (text)} 
The temporal logic specification is a required input of \tool.
A specification $\phi$ is given as a string by using following notation $\levent \rightarrow \{\text{``F",} ``<>"\}$, $\lalways \rightarrow \{\text{``G"}, ``[]"\}$, and $\luntil \rightarrow \text{``U"}$, 
for the \emph{eventually}, \emph{always}, and \emph{until} temporal operators, respectively.
The logical operators are given by $\neg \rightarrow \{ ``!", ``\sim"\}$,
$\land \rightarrow \{\text{``\&\&", ``\&"}\}$ and $\lor \rightarrow \{\text{``\textbar\textbar", ``\textbar"}\}$ for \emph{negation}, \emph{conjunction} and \emph{disjunction}, respectively.
One example of writing an STL specification $\phi = (\levent_{[0,4]} s>2)  \land (\lalways_{[2,4]} s\leq4)$ as an input for \tool\, is the following $``(\text{F}[0,4] s>2) \&\& (\text{G}[2,4] s<=4)"$.
Note that the predicates differ in MTL from the STL notation, e.g., $``\text{F}[0,4] RegionA"$. 
However, for wSTL specifications, there are some additions to the notation for the temporal and logical operators.
For instance, let us consider $\phi = \land^p(\lalways^{w_1}(s>2), \levent^{w_2} s<1)$ the corresponding input should be written as $``\&\&^\wedge p(\text{G}^\wedge \text{w1} (s>2), \text{F}^\wedge \text{w2} (s<1))"$, where $p$, $w_1$, and $w_2$ are the corresponding logical and temporal weights defining the preferences and importance, respectively, and given as Python dictionaries.

\subsection{Parser}
We parse the input specification using a well-known and well-maintained parser generator, ANTLR4~\cite{parr2013definitive}. 
One advantage of using ANTLR4 to parse specifications is that it generates grammar files for multiple programming languages such as Java, C\#, Python 2 and 3, Go, C++, Swift, JavaScript, and TypeScript.
This allows users to adapt \tool\, to their desired or preferred programming language.
For \tool\, we use ANTLR4 to generate grammar files 
($\texttt{stl.g4}$, $\texttt{mtl.g4}$, $\texttt{wstl.g4}$) for Python 3.

\subsection{Abstract Syntax Tree (AST)} 
\begin{figure}
    \centering
    \includegraphics[width=0.55\linewidth]{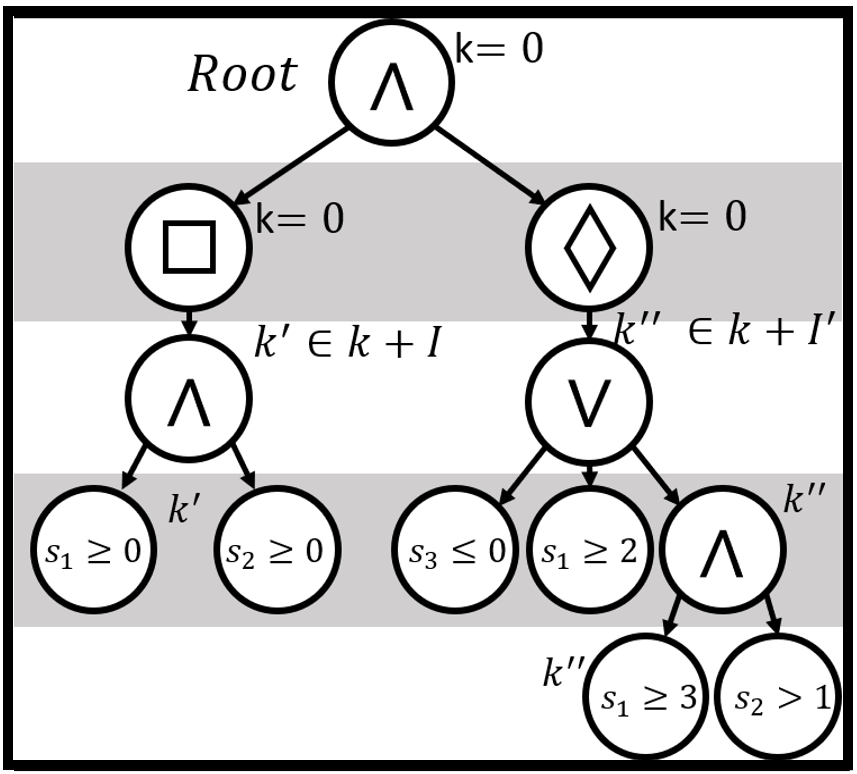}
    \caption{Abstract Syntax Tree of STL specification $\phi_{ex}$.}
    \label{fig:ast}
    \vspace{-12pt}
\end{figure}

We create Abstract Syntax Trees (AST) from grammar files that capture the specifications' structure.
Depending on the temporal logic language used, the user might choose among the different scripts $\texttt{stl.py, mtl.py, wstl.py}$ for STL, MTL, and wSTL, respectively.
An example of how an AST abstracts an STL specification $\phi_{ex}$ is given in Fig. \ref{fig:ast}.

\begin{equation*}
    \begin{aligned}
    \phi_{ex} \coloneqq & \lalways_I ((s_1 \geq 0) \land (s_2 \geq 0) ) \land \\
                        & \, \levent_{I'} ((s_3 \leq 0) \lor (s_1 \geq 2) \lor ((s_1 \geq 3) \land (s_2>1))),
    \end{aligned}
\end{equation*}

where $I$ and $I'$ are intervals of time and $s_1,\, s_2, \, \text{and } s_3$ are signals.

Note that in an AST, leaf nodes correspond to predicates, whereas the root and intermediate nodes correspond to temporal and logical operators.
In the particular case of wSTL, weights are set in the links connecting the nodes.
Users can also modify the symbols capturing the semantics of the temporal logic language and generate their own grammar files.

\begin{remark}
We choose an AST as an intermediate representation because it is a very convenient way to abstract, manipulate, and analyze temporal logic specifications for use within various frameworks such as reinforcement learning (RL), model predictive control (MPC), and non-linear optimization (NLO).
\end{remark}

\subsection{General Methods}
Among the general methods implemented in $\texttt{stl.py}$, $\texttt{mtl.py}$, $\texttt{wstl.py}$,
transforming a specification into PNF form and negating a formula while pushing \emph{not} operators down to leaves are very useful for further processing.
Another common and important method is the computation of the specification's time horizon or bound~\cite{stl-horizon}.
It is useful for synthesis and planning, e.g., MPC~\cite{yu2022model}.
Finally, we have four different methods for computing robustness based on traditional~\cite{donze2012temporal} and AGM~\cite{mehdipour2019arithmetic} robustness for simple traces and trace batches.
These are useful for many frameworks, such as inference of temporal logic specifications~\cite{bombara2016decision}. 

\subsection{AST to MILP}
A subsequent advantage of having the temporal logic specification as an AST is that it enables recursive implementation of the MILP encoding over the tree.
Specifically, for each node, we use the variables from the children (i.e., subformulas) to encode the operation associated with the node. 
The parent then uses the node's satisfaction variable.
For leaves (nodes without children), the encoding is direct.
At the root, we enforce its satisfaction to ensure that the specification holds.
For this, the user has to choose the script among $\texttt{stl2milp.py, mtl2milp.py, wstl2milp.py}$ depending if the encoding corresponds to STL~\cite{sadraddini2015robust, raman2014model}, MTL, or wSTL~\cite{cardona2023wstl}, respectively.
The output of this step is a Gurobi model $\texttt{GRB.model()}$ that can be extended with additional constraints (e.g., capturing dynamics) and cost functions.

\subsection{Dynamics}
Note that the temporal logic specification is added to the optimization problem in the previous block. 
The goal is to maximize the robustness, guaranteeing satisfaction according to the soundness property if positive.
However, depending on the application, a specification might not be the only constraint governing the problem; the problem of interest may contain dynamics governing the signal generators.
Dynamics can be defined using recurrence equations, graph-based models such as transition systems, Markov Decision Processes (MDP), or any other that the user might prefer as long as the constraints are linear.
For graph-based models below, we used LOMAP~\cite{ulusoy2013optimality, kamale2021automata}.
These constraints must be added via
$\texttt{GRB.model.addConstrs()}$, 
an example of how to connect the dynamics to the variables from the specification is provided in Sec.~\ref{sec:control-synthesis}.

\subsection{MILP-Solver}
A primary motivation for the modular design of \tool\, is to allow users to easily augment or replace blocks in the toolflow.
Thus, any solver capable of solving MILP problems could be inserted in this block.
Due to the available academic licenses and fast performance, we have chosen Gurobi~\cite{gurobi} as the primary solver.
Once the temporal logic specification and dynamics are added to the Gurobi model as variables and constraints, the user can set the objective functions according to the prescribed performance metrics, e.g., robustness with respect to the specification, control effort, state deviation from nominal values, etc.

\subsection{Solution}
The last block corresponds to the output whose format depends directly on the data structure of the dynamics being imposed in the model.

\section{Capabilities of \tool}
In this section, we demonstrate some of the capabilities and case studies of previous work implemented using \tool.
The first capability we consider is the parsing for STL, MTL, and wSTL specifications.
Syntactic sugar or language extensions of the current temporal logic frameworks are straightforward to implement due to the parsing method that \tool\, uses.
Additionally, as \tool\, keeps the parsing and synthesis separated, parsing the specification can be used for multiple purposes.

We start by examining some specific capabilities of \tool, such as trajectory optimization and control synthesis.

\subsection{Trajectory optimization}
\label{sec:trajectory-optimization}
Trajectory optimization is performed when a temporal logic specification is defined, and no dynamics are imposed.
The only requirement to solve this problem is the state or variable (predicate) bounds.
A simple example could be the following:
\begin{example}[Trajectory optimization example]
Consider two signals $s_1$ and $s_2$, and the following STL and wSTL specifications
\begin{equation}
    \begin{aligned}
    \phi_{STL} =& \left ( \lalways_{[1,5]} (s_1\geq 7) \lor \lalways_{[1,5]} (s_2 \leq 2) \right ) \land \\
    &\left ( \levent_{[5,10]} (s_1 \leq 3) \lor \levent_{[5,10]}(s_2 \geq 8)  \right ),\\
    \phi_{wSTL} =& \land^{p_1} \left ( \lor^{p_2} (\lalways^{w_1}_{[1,5]} (s_1\geq 7), \lalways^{w_2}_{[1,5]} (s_2 \leq 2)) \right . ,\\
    & \left . \lor^{p_2} (\levent^{w_1}_{[5,10]} (s_1 \leq 3), \levent^{w_2}_{[5,10]}(s_2 \geq 8))\right ),
    \end{aligned}
\end{equation}
with  variable bounds $0 \leq s_1 \leq 10$, $0 \leq s_2 \leq 10$, 
initial conditions $s_1(0) = 0$ and $s_2(0) = 0$, and logical operators weights $p_1=[0.5, 0.5]$, $p_2=[0.2, 0.8]$ which defines the preferences of the subformula to be satisfied, and $w_1=[1, 2, 3, 4, 5]\times 10^{-1}$, $w_2=[5, 4, 3, 2, 1]\times 10^{-1}$ the temporal weights defining the importance of time-steps for satisfaction.
\end{example}

\begin{figure}[t]
    \centering
    \subfloat[$\phi_{STL}$ solution over time.]{
    \includegraphics[width=.8\linewidth]{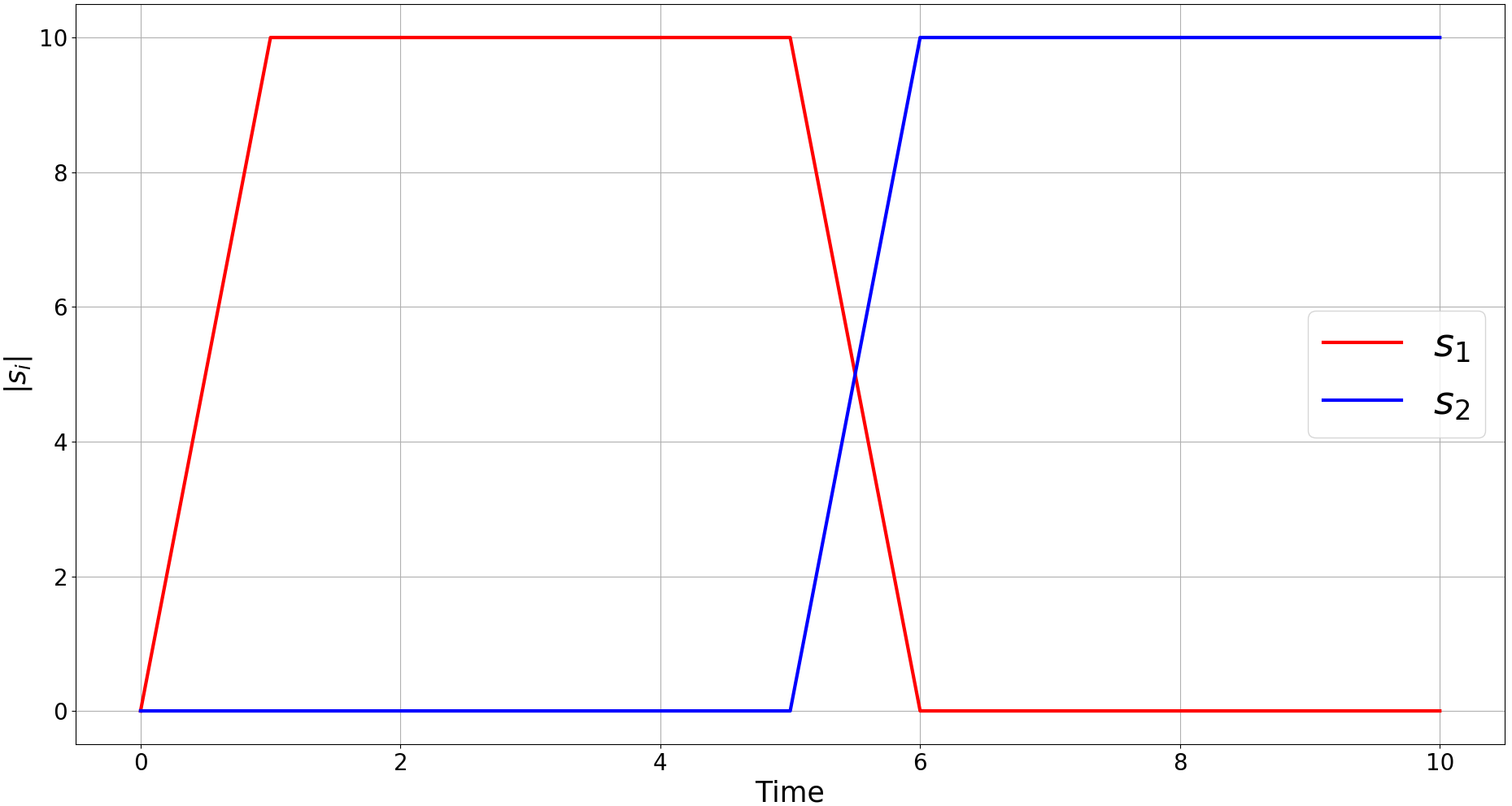}
    \label{fig: stl_example}
    }\hfill
    \subfloat[$\phi_{wSTL}$ solution over time.]{
    \includegraphics[width=0.8\linewidth]{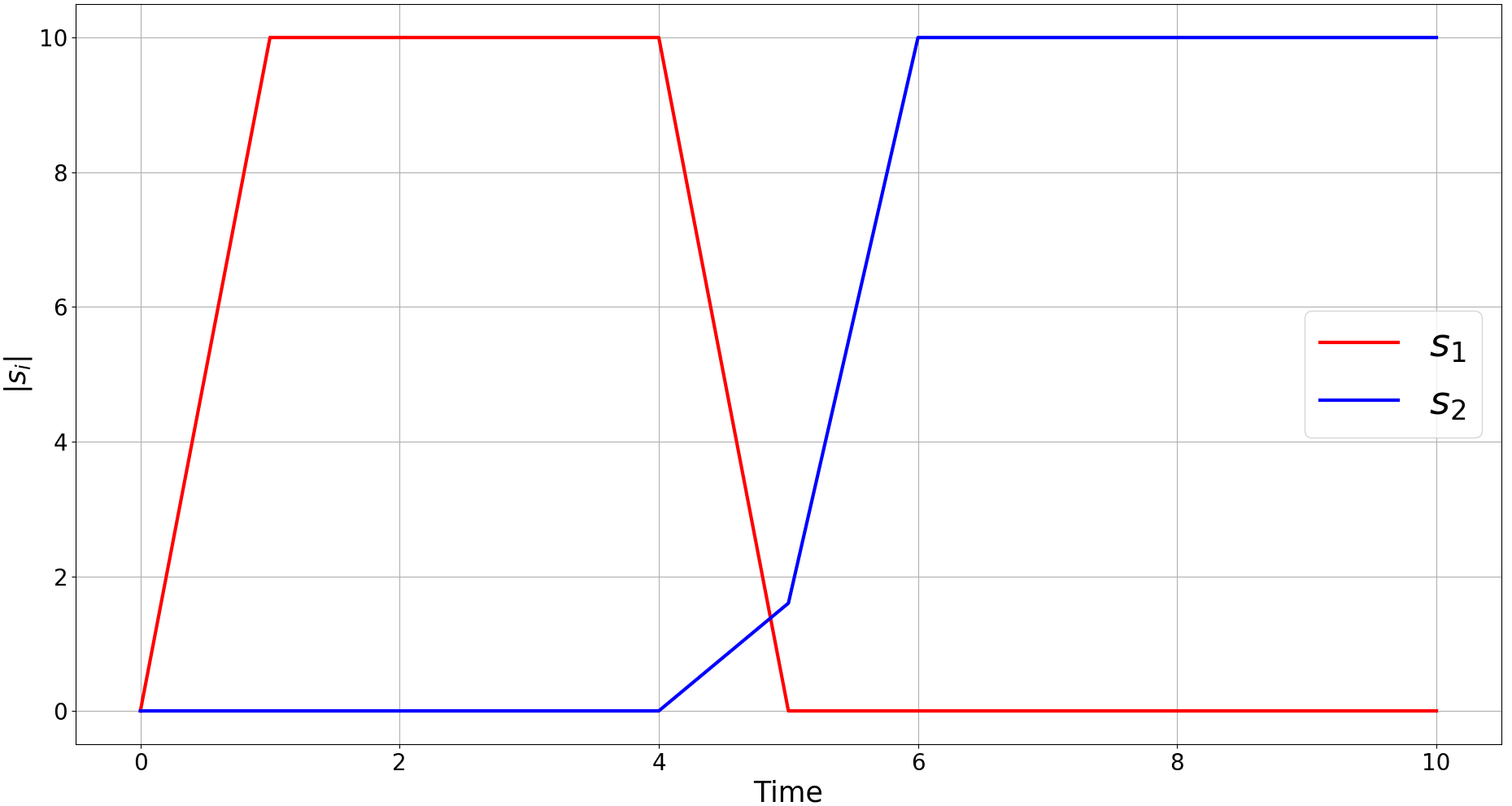}
    \label{fig: wstl_example}
    }
   \caption{Trajectory generated with STL and wSTL without dynamics imposed.}
    \label{fig:example-stl-wstl}
\end{figure}

In Fig.~\ref{fig:example-stl-wstl}, we show the trajectories synthesized by \tool\, under the previous conditions. 
Due to space limitation, here we show how to use \tool\, for solving the STL example. 
However, wSTL uses a similar workflow to STL.
First, we use the method 
$\texttt{stl\_synth\_ctrl(formula, vars\_ub, vars\_lb)}$ where $\texttt{formula}$ is the specification, $\texttt{vars\_ub}$ and $\texttt{vars\_lb}$ are the upper and lower bounds of the variables.
Next, we create the Gurobi object 
\begin{minted}{python}
stl_milp = stl2milp(ast, robust=True)
\end{minted}
using $\texttt{stl2milp.py}$, which the AST as input, and we define robustness to \mintinline{python}/True/ to maximize robustness. 
Otherwise, only satisfaction with the specification is taken into account.
Then, to actually translate the AST into the MILP, we use the following method \mintinline{python}/stl_milp.translate(satisfaction=True)/, in which we impose satisfaction to be \mintinline{python}/True/ so that if the specification is infeasible, Gurobi returns model infeasible.
Lastly, we use the following Gurobi method $\texttt{stl\_milp.model.optimize()}$ to run the optimization.

\subsection{Open-loop control synthesis for linear time-invariant systems}
\label{sec:control-synthesis}
Recall that the specification can be solved with and without considering dynamics, which basically defines the difference between finding an optimized trajectory or synthesizing a controller.

Consider the following linear time-invariant system:
\begin{equation}
\label{eq:dynamics}
    \begin{aligned}
        s({k+1}) &= As(k) + B u(k) + D,\\
        y(k) &= Cs(k),
    \end{aligned}
\end{equation}
where $s(k) \in \mathbb{R}$ is the state variable at time $k \in \mathbb{Z}_{\geq 0}$, $u$ is the input of the system, $y$ is the output, $A$, $B$, $C$, and $D$ are the state transition, input, and drift matrices of appropriate sizes.
In addition to the dynamic constraints, control saturation constraints of the form $\|u(k)\|_1 \leq 1$, $\|u(k)\|_\infty \leq 1$ can also be considered.

Imposing dynamics constraints might imply that the user wants to modify the optimization function and not just focus on maximizing the specification's robustness.
One example of the optimization function that leverages multiple behaviors is the following
\begin{equation}
\label{eq:cost-function}
    J = \lambda \rho(s,\phi,0) - \alpha^\top \|s\|_1 - \beta^\top \|u\|_1,
\end{equation}
where $\rho(s,\phi,0)$ is the robustness of the specifications, $s$ is the dynamics variables  that are coupled to the predicate variables, and
$\lambda \geq 0$, $\alpha \geq 0$, and $\beta \geq 0$ are regularization weights.
Note that maximizing the cost function and setting $\lambda=0$, $\alpha=0$, and $\beta=0$ implies the feasibility case, so the satisfaction of the specification is considered while constrained to the dynamics.
When $\lambda>0$, $\alpha=0$, and $\beta=0$, the goal is to maximize the robustness and is equivalent to the trajectory optimization described in Sec.~\ref{sec:trajectory-optimization}.
On the other hand, if we allow $\lambda>0$, $\alpha>0$, and $\beta>0$, then the objective includes the satisfaction of the specification and maximization of the robustness while minimizing both the deviation from 0 of the signals and the use of control signals.
The value difference between $\lambda$, $\alpha$, and $\beta$ defines the regularization and priorities of the user.

\begin{figure}[t]
    \centering
    \subfloat[Evolution of $s_1$ with time.]{
    \includegraphics[width=0.8\linewidth]{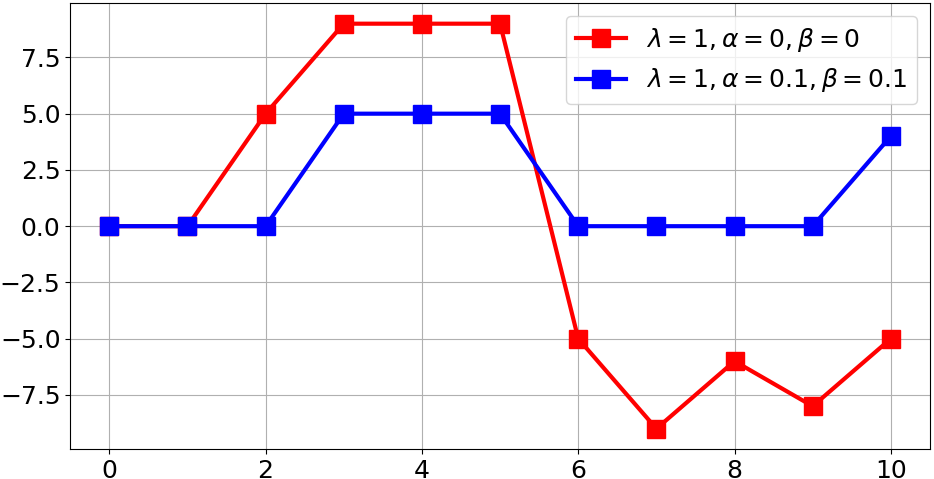}
    \label{fig: x_vs_t}
    }\hfill
    \subfloat[Evolution of $s_2$ with time.]{
    \includegraphics[width=0.8\linewidth]{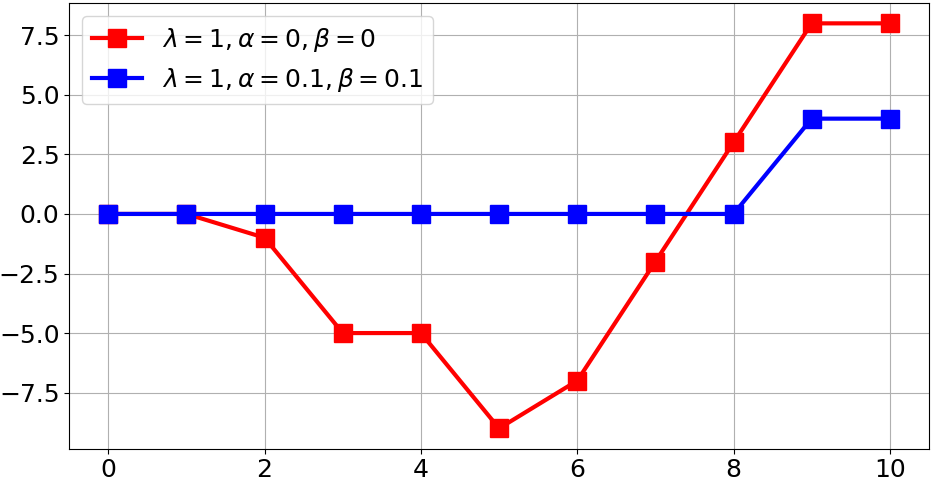}
    \label{fig: y_vs_t}
    }\hfill
    \subfloat[Evolution of $u_1$ with time.]{
    \includegraphics[width=0.8\linewidth]{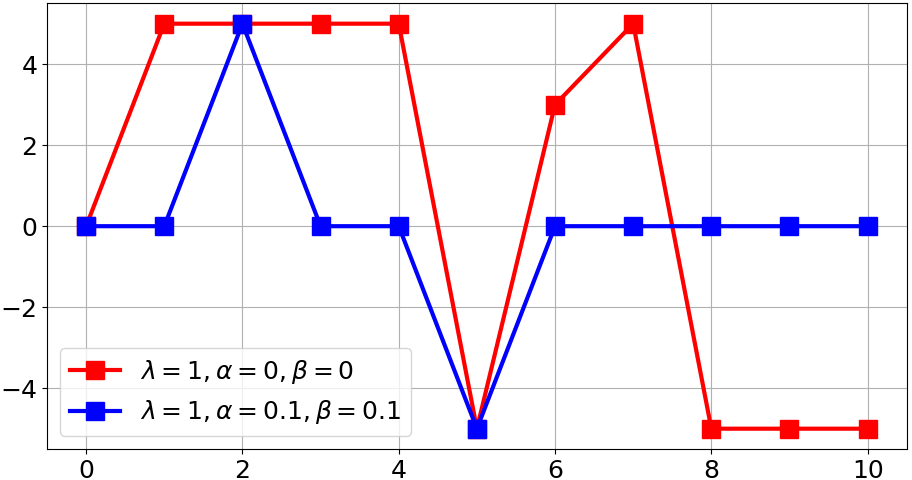}
    \label{fig: u_vs_t}
    }\hfill
    \subfloat[Evolution of $u_2$ with time.]{
    \includegraphics[width=0.8\linewidth]{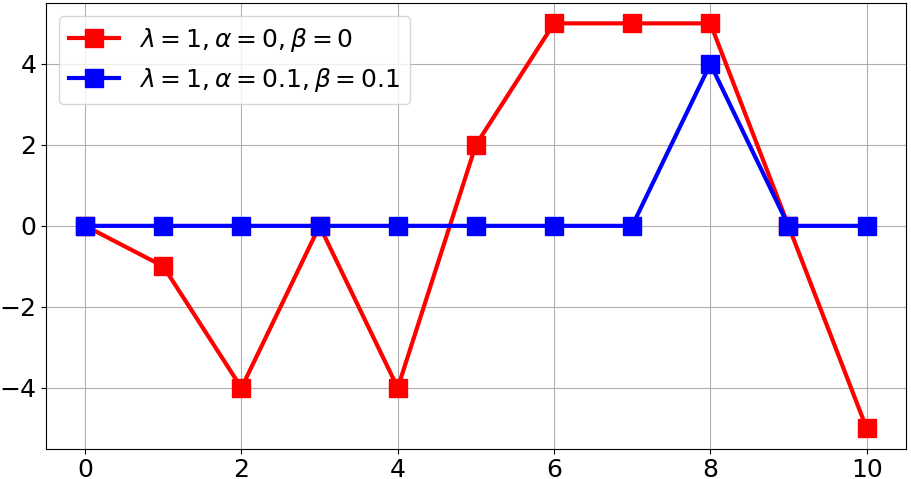}
    \label{fig: v_vs_t}
    }
   \caption{Control synthesis for satisfaction of $\phi_{cs}$ coupled with dynamic constraints and control saturation. 
 Red lines maximize only robustness, while blue lines maximize robustness while minimizing control signal generation and signal deviation.}
    \label{fig:synthesis-control}
\end{figure}

\begin{example}
Let us consider a linear dynamic system as in \eqref{eq:dynamics}, an optimization function of the form given in \eqref{eq:cost-function}, and a simple STL specification $\phi_{cs} = (\lalways_{[3, 5]} (s_1 >= 3)) \land (\lalways_{[9,10]} (s_2 >= 2))$.
With
$$ A=\begin{bmatrix}
1 & 1 \\
0 & 1\\
\end{bmatrix}, 
B = \begin{bmatrix}
1 & 0 \\
0 & 1 \\
\end{bmatrix},
C=[1\; 0],  
D=[0, 0]^\top $$
variable bounds $-9 \leq s_1 \leq 9$, $-9 \leq s_2 \leq 9$, and control bounds $-5 \leq s_1 \leq 5$ and $-5 \leq s_2 \leq 5$. 
\end{example}
In Fig. \ref{fig:synthesis-control}, we show two different examples of the regularization parameters. 
One considers only robustness maximization (red lines), and the other maximizes robustness and minimizes the use of the control signal and deviation of the variables from zero (blue lines).
It is notable how the inclusion of minimization of control signal usage of $u_1$ and $u_2$ and deviation from zero for the signals $s_1$ and $s_2$ modifies the solution and control signal generation in comparison to when just robustness is chosen. 
It is worth mentioning that even though the generated trajectories are different, both satisfy the specification.

In terms of how to define this problem, using \tool\, requires the creation of a Gurobi object and translating the specification into an AST and later into the MILP, as in the trajectory optimization in Sec. \ref{sec:trajectory-optimization}.
The first modification involves coupling the variables from the predicate to the variables we want to use for defining the dynamic constraints. 
For this, we first create all the necessary Gurobi variables through the time horizon. 
The following code snippet shows an example of how variable $s_1(k)$ is created in the Gurobi model. The variable $s_2(k)$ can be created the same way.
\begin{minted}[breaklines, breakanywhere]{python}
for k in range(time_horizon):
    name = "s1_{}".format(k)
    s1[k] = stl_milp.model.addVar(vtype=grb.GRB.CONTINUOUS, lb=vars_lb, ub=vars_ub, name=name)
\end{minted}
Then it is required to indicate which variables are related, which is achieved by a simple command such as \mintinline{python}/stl_milp.variables['s1'] = s1/.
Next, by defining all the required variables as shown before, we can define the dynamic constraints as follows

\begin{minted}[breaklines, breakanywhere,breakautoindent=false]{python}
for k in range(time_horizon-1):
    stl_milp.model.addConstr(s1[k+1] == A[0][0]*s1[k] + A[0][1]*s2[k] + B[0][0]*u1[k])
    stl_milp.model.addConstr(s2[k+1] == A[1][0]*s2[k] + A[1][1]*s2[k] + B[0][1]*u2[k])
\end{minted}

An additional set of constraints that must be defined are the initial conditions of the variables, which can be imposed as follows $$\texttt{stl\_milp.model.addConstr(s1[0] == 0)}.$$
The final step requires redefining the objective function. 
For this, we need to add auxiliary variables to capture the absolute value as follows

\begin{minted}[breaklines, breakanywhere,breakautoindent=false]{python}
stl_milp.model.addConstrs(s1_aux[k] == grb.abs_(s1[k]) for k in range(time_horizon)) 
stl_milp.model.addConstrs(s2_aux[k] == grb.abs_(s2[k]) for k in range(time_horizon))
\end{minted}

Then we can define a state cost to capture $\|s(k)\|_1$  
as follows
\begin{minted}[breaklines]{python}
state_cost=sum(s1_aux[k] + s2_aux[k] for k in range(time_horizon))
\end{minted}
Then the addition of this term to the objective function can be added as follows

\begin{minted}[breaklines, breakanywhere]{python}
stl_milp.model.setObjectiveN(state_cost, 2, weight=alpha, name='state_cost')
\end{minted}
The same procedure can be used for $\|u(k)\|_1$. Due to space limitations, we avoid encoding it here explicitly. The last step will require optimizing the updated model with $\texttt{stl\_milp.model.optimize()}$.

\subsection{Extended capabilities of \tool}
Next, we highlight some relevant examples of adapting or extending \tool\, to other problem domains in previously published work.
\begin{figure}[t]
    \centering
    \subfloat[Partial Satisfaction.]{
    \includegraphics[width=.6\linewidth]{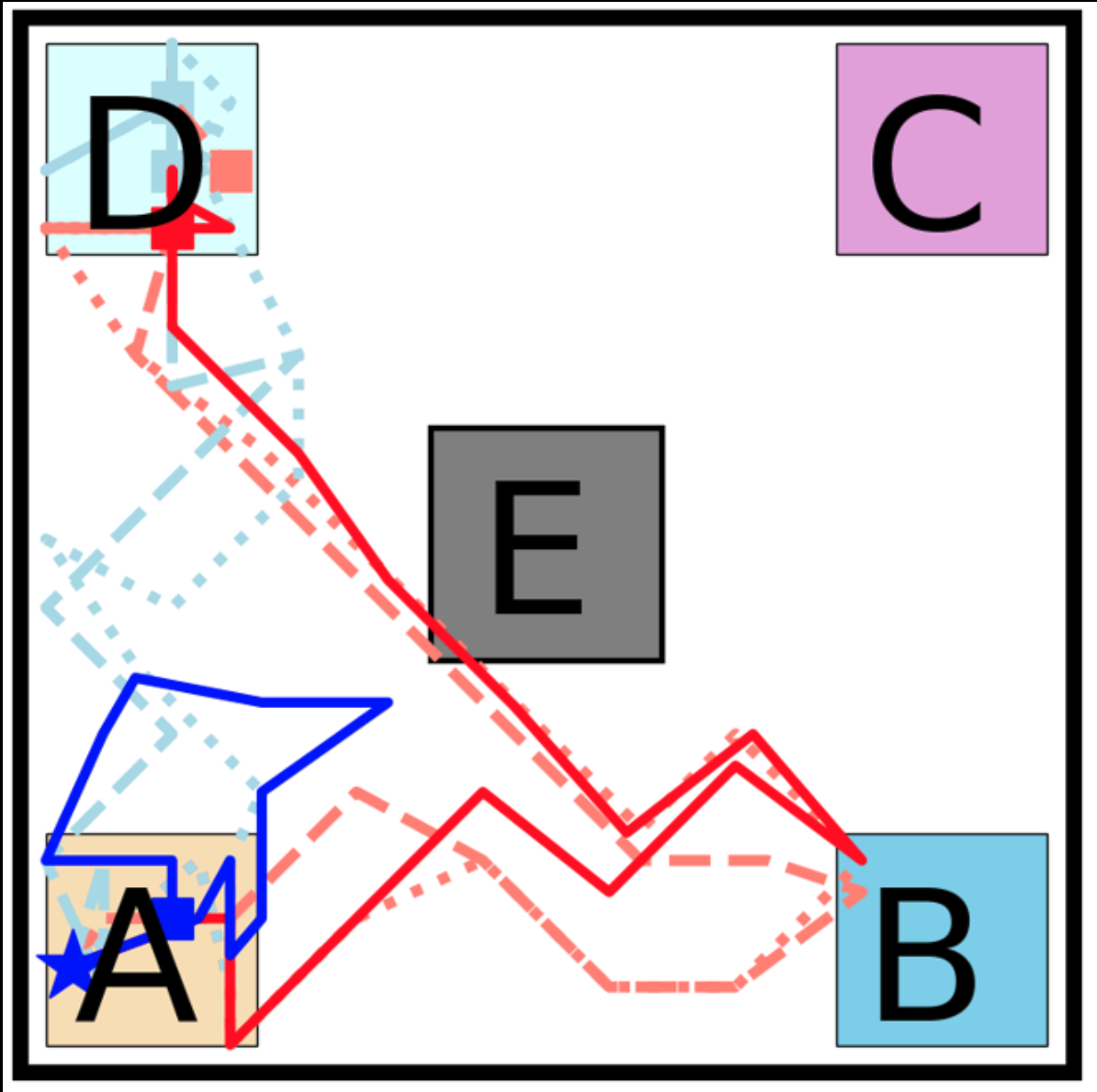}
    \label{fig: ps}
    }\hfill
    \subfloat[Planning for swarms with splitting and merging]{
    \includegraphics[width=0.8\linewidth]{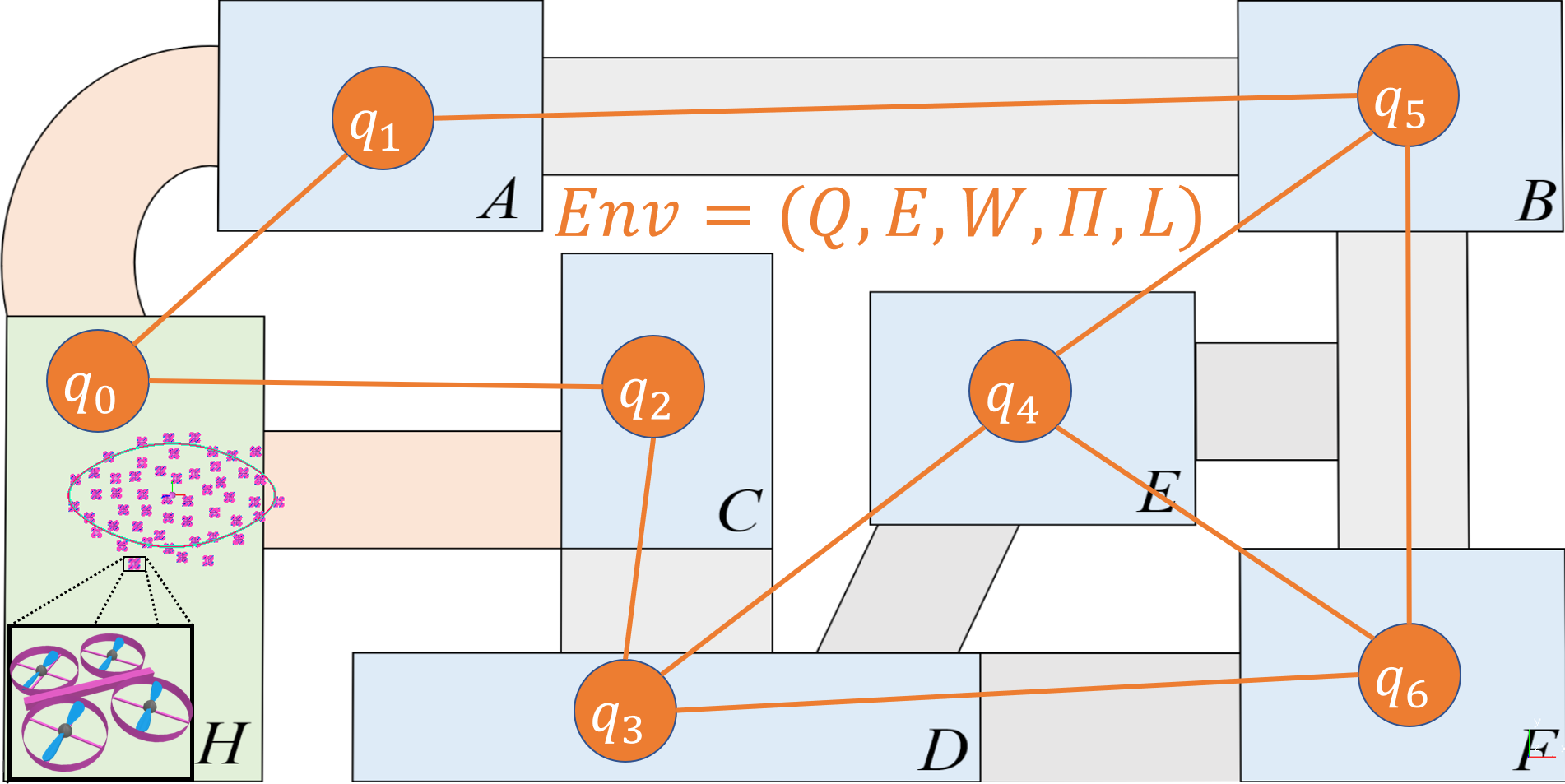}
    \label{fig: swarms}
    }\hfill
    \subfloat[Planning for modular robotics]{
    \includegraphics[width=0.8\linewidth]{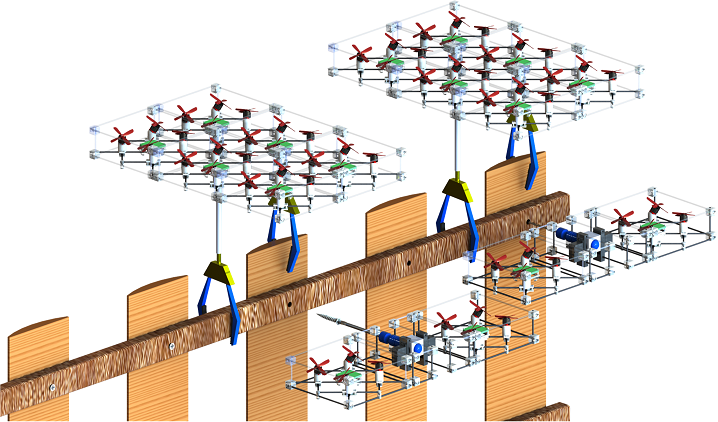}
    \label{fig: modular}
    }\hfill
    \subfloat[Capability Temporal Logic implementation]{
    \includegraphics[width=0.8\linewidth]{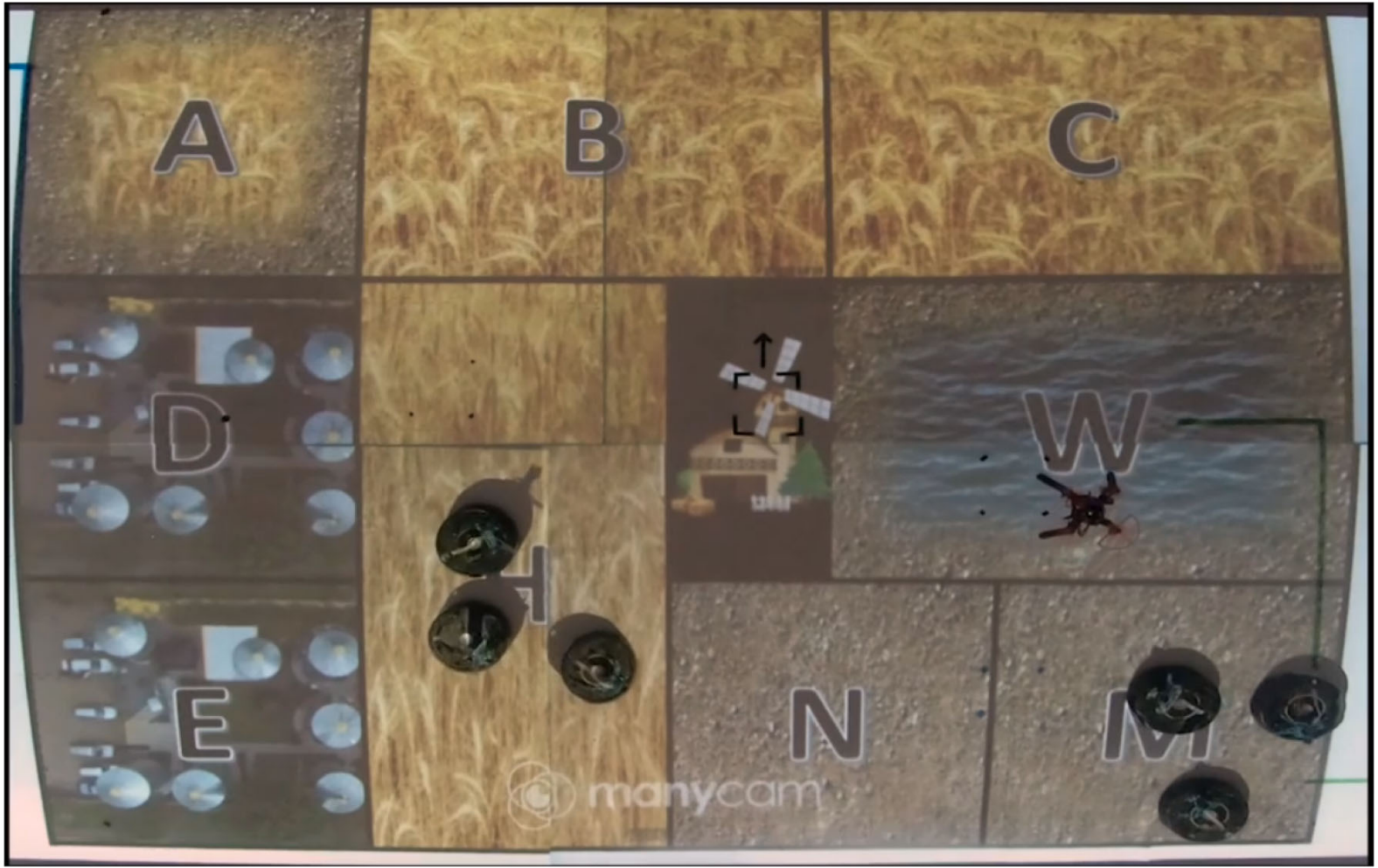}
    \label{fig: catl}
    }
   \caption{Extensions and applications using \tool.}
    \label{fig:case_studies}
\end{figure}

\subsubsection{Satisfaction semantics encoding}

\emph{Partial satisfaction (PS) of STL specifications:}

In certain cases, users may wish to modify the semantics of STL or another logic for a specific use case.
One such example is when a specification is infeasible or conflicting.

In such cases,
the solver returns an ``infeasible model'' as the output, 
even though partial satisfaction may be acceptable.
Based on this idea in \cite{cardona2023partial}, we define an encoding that accounts for the case of infeasible or conflicting specifications, satisfying them as much as possible.
The encoding is inspired by the notion of fractions of satisfactions instead of the standard Boolean representation.
These modified semantics can be captured by
replacing the $\texttt{stl2milp.py}$ with the encoding in \cite{cardona2023partial}, 
allowing
\tool\, to plan for partially satisfiable specifications. 
The results of this extension are shown in Fig. \ref{fig:case_studies}(a), where an agent is tasked to visit regions $D$ and $B$ simultaneously, and the encoding decides to partially satisfy the specification by going to at least one of the places.
Standard STL encodings could not satisfy any of the tasks, so it minimally violated both.

We extend this work to also capture user preferences and importance priorities for Boolean and temporal operators.
These preferences are captured by extending partial satisfaction on weighted signal temporal logic (wSTL) specifications~\cite{cardona2023preferences}. 
Additionally, we allow the use of Boolean and temporal \emph{inclusive} and \emph{exclusive} operators. 

\subsubsection{Predicate semantics}

\emph{Planning for swarms of robots:}

Changing the main temporal logic encoding is not the only extension that can be performed with \tool.
In \cite{cardona2023swarms}, we modify the semantics of the predicates in the MTL encoding $\texttt{mtl2milp.py}$, to be satisfied if a swarm is at the state and a time step is required. 
The main objective is to find a high-level plan for a swarm that can split and merge to satisfy temporal and logical tasks.
The environment is abstracted as a transition system, and the constraints that need to be added to the $\texttt{GRB.model()}$ are inspired by the flow network problem.
We modify the objective function to 
force swarms to split only when needed for the satisfaction of the mission.
A simple example is shown in Fig.~\ref{fig:case_studies}(b), where the transition system is shown, and there is a swarm of quadrotors waiting for the high-level commands.

\emph{Planning for modular robots:}

A more complex case in terms of the dynamics is considered in \cite{cardona2022planning}, where the goal is to plan for modular aerial robots and modular tools to satisfy tasks that require tools in the states of a transition system.
The specification is captured using an MTL encoding and, once again, modifying the semantics of the predicates in $\texttt{mtl.py}$ and $\texttt{mtl2milp.py}$.
These are satisfied only if there is a modular robot in a configuration and tools that can satisfy the task for the duration of time requested and where and when the specification requires it.
We impose extra constraints to guarantee the existence of modular robots in just one state and configuration at a time.
We add three different objective functions.
First, it captures the energy spent when a modular robot goes from one configuration to another.
Second, it captures the energy of traversing the environment in a particular configuration.
Lastly, it considers the performance of satisfying a task in a specific configuration.
An example of the modular robots that coordinate to build a fence is shown in Fig.~\ref{fig:case_studies}(c).

\subsubsection{Domain-specific language fragment of STL} 

\emph{Capability temporal logic (CaTL):}

A fragment of STL was introduced to define a richer specification language in \cite{leahy2021scalable, jones2022scratchs}, called Capability Temporal Logic (CaTL).
The main idea here is that CaTL accounts for the use of different agent classes and the quantity of them requested in a region for a duration of time.
This is an important example extended from \tool\, because it involved the creation of new grammar files $\texttt{catl.g4}$ to recognize the semantics of CaTL specifications.
But as it is a fragment of STL, instead of creating an encoding to go from CaTL's AST to MILP, what is performed is a translation from CaTL to STL and further STL to MILP.
This shows the advantages of having a modular and versatile architecture in \tool.
An example of the kind of missions that can be satisfied with CaTL is shown in Fig. \ref{fig:case_studies}(d), where robots with different capabilities satisfy tasks in a physical environment.


\section{Conclusions}
We presented \tool, a Python-based software to support temporal logic frameworks such as STL, wSTL, and MTL for synthesizing controllers with and without dynamic constraints.
We describe the architecture of \tool\, on how we generate grammar files using ANTLR later to create an AST representation of the given specification.
This AST abstraction allows the user to use \tool\, for other frameworks such as reinforcement learning, machine learning, and non-linear optimization.
Thanks to the modularity and architecture of \tool, more temporal languages or extensions and fragments of the existing ones can be easily implemented.
We show we can use general methods such as transformations, time-horizon, and robustness computation or control synthesis by formulating a MILP that is solved using Gurobi.
However, it can be modified and extended to the user's preferred solver.
We show how dynamics can be added to \tool\, and how different objective functions define the desired behavior of the computed solution, such as satisfiability, robustness, control input usage minimization, and minimal signal deviation from zero.
The modularity and extensibility capability in \tool\, allows it to be easily modified and extended by the user to customize it to the application. 
We show examples of previously published work using \tool\, from domains such as swarms, modular robots, and robots with heterogeneous capabilities.

\bibliographystyle{ieeetr}
\bibliography{references.bib}
\end{document}